\documentclass{iau}

\usepackage{amsmath}
\usepackage{graphicx}
\usepackage{upgreek}

\begin{document}

\lefttitle{R. M. Rich, B. Thorsbro \& C. I. Johnson}
\righttitle{Is the Nuclear Star Cluster a Bulge Fossil Fragment?}

\jnlPage{1}{6}
\jnlDoiYr{2026}
\doival{10.1017/xxxxx}

\aopheadtitle{Proceedings IAU Symposium}
\editors{M. Zaja\v{c}ek, T. Je\v{r}\'{a}bkov\'{a}, V. Karas, R. Sch\"odel \& P. Sukov\'{a}, eds.}

\title{Is the Nuclear Star Cluster a Bulge Fossil Fragment?}

\author{%
R. Michael Rich$^{1}$,
Brian Thorsbro$^{2,3}$,
and Christian I. Johnson$^{4}$%
}

\affiliation{%
$^{1}$Department of Physics and Astronomy, University of California,
Los Angeles, CA 90095-1547, USA \\
$^{2}$Division of Astrophysics, Department of Physics, Lund University,
Box 118, SE-22100 Lund, Sweden \\
$^{3}$Universit\'{e} C\^ote d'Azur, Observatoire de la C\^ote d'Azur,
CNRS, Laboratoire Lagrange, 06000 Nice, France \\
$^{4}$Space Telescope Science Institute, 3700 San Martin Drive,
Baltimore, MD 21218, USA%
}

\begin{abstract}
The Milky Way nuclear star cluster (NSC) is unlike a conventional globular cluster: it contains a broad metallicity distribution and stellar populations formed over an extended interval, while its detailed abundances show a close connection to the inner Galactic bulge.  We ask whether the NSC should instead be compared with the proposed ``bulge fossil fragments'' Terzan~5 and Liller~1, complex stellar systems interpreted as surviving remnants of massive structures involved in early bulge assembly.  We place this comparison against the field-bulge metallicity distribution measured by the Blanco DECam Bulge Survey and against high-resolution infrared abundance measurements in the NSC.  A broad metallicity distribution by itself is not diagnostic, because the bulge field is intrinsically broad and strongly structured.  More discriminating tests are the presence of chemically narrow and age-coherent subpopulations, the relation between [$\upalpha$/Fe] and [Fe/H], light-element abundance patterns, spatial segregation, and chemo-dynamical differences among populations.  Existing NSC data establish strong chemical links with the bulge but do not support a simple one-to-one analogue of either Terzan~5 or Liller~1.  In particular, metal-rich $\upalpha$-enhanced stars in the NSC provide an important contrast with the near-solar [$\upalpha$/Fe] metal-rich populations of the best-studied fossil-fragment candidates.  We argue that the BFF hypothesis is most useful as a falsifiable framework for identifying an early bulge-building component within a composite NSC assembled through in-situ star formation, migration, and cluster infall.
\end{abstract}


\maketitle

\vspace{-0.4cm}
\section{Introduction}

How did the Milky Way acquire its nucleus?  The answer must account simultaneously for the nuclear star cluster (NSC), the surrounding nuclear stellar disc and bulge, and the supermassive black hole Sgr~A*.  The NSC is particularly difficult to interpret because extreme and spatially variable extinction and severe crowding limit direct age measurements and make high-resolution spectroscopy observationally expensive.  Nevertheless, the emerging population picture is already sufficient to show that the NSC is not well described as an ordinary massive globular cluster.

The NSC contains stars spanning a wide range of ages, from recent star formation to an old population, and its red giants cover a broad metallicity distribution \citep{pfuhl11,rich17}.  High-resolution Keck/NIRSPEC spectroscopy found a median [Fe/H] near $-0.16$ dex in an early sample of old NSC giants, with stars extending from roughly [Fe/H]$\sim-1$ to above $+0.5$ dex; the resulting metallicity distribution resembled the Galactic bulge more closely than the local disc or halo \citep{rich17}.  Subsequent detailed abundance work found a metal-rich population with enhanced [Si/Fe] relative to local thin-disc stars \citep{thorsbro20}, while larger recent infrared samples reinforce the connection between NSC chemistry and that of the inner bulge \citep{ryde25,nandakumar25}.

A broad [Fe/H] distribution and an extended star-formation history are therefore central characteristics of the NSC.  They are also the features that make Terzan~5 and Liller~1 particularly interesting comparison systems.  These objects were historically classified as globular clusters, but their multiple iron-abundance and age populations instead motivated the idea that they are surviving remnants of much more massive structures that participated in the early assembly of the Galactic bulge: ``bulge fossil fragments'' (BFFs) \citep{ferraro21}.

Here we ask a deliberately comparative question: does the NSC preserve the population signatures expected of a BFF, or is its chemical complexity better understood as the superposition of ordinary bulge stars, genuine globular-cluster debris, and repeated in-situ nuclear star formation?  We do not require that the present NSC be the intact remnant of a single BFF that migrated to the Galactic center.  A BFF-like early component could instead be one ingredient of a composite nucleus.  This formulation turns the analogy into a set of observational tests rather than an origin story that must be accepted or rejected as a whole.

\section{What does a bulge fossil fragment look like?}

Terzan~5 provides the clearest chemical benchmark.  High-resolution spectroscopy established distinct iron-abundance populations and different $\upalpha$-element behaviour: the subsolar component is $\upalpha$-enhanced, while the metal-rich component has approximately solar [$\upalpha$/Fe], consistent with enrichment on a longer timescale that included Type~Ia supernovae \citep{origlia11}.  Its metallicity distribution is demonstrably multi-modal rather than that of a monometallic globular cluster \citep{massari14}.  Photometry subsequently showed distinct main-sequence turnoffs, with the principal subsolar population being old and the supersolar population several Gyr younger \citep{ferraro16}.  New JWST photometry now resolves the turnoff region particularly clearly, giving ages of $12.5\pm0.5$ Gyr and $4.7\pm0.5$ Gyr for the two principal components, with evidence for still younger stars \citep{zullo26}.

Liller~1 exhibits the same qualitative combination of an old population and a much younger component.  Deep photometry identifies an old population near 12 Gyr and a young component formed only a few Gyr ago, motivating its inclusion with Terzan~5 in the BFF class \citep{ferraro21}.  Recent high-resolution spectroscopy strengthens the analogy: Liller~1 contains chemically distinct subsolar and supersolar populations, but its old component does not show the Na--O anticorrelation characteristic of genuine globular clusters \citep{alvarez24}.  Terzan~5 likewise lacks the light-element anticorrelation expected for an ordinary globular cluster in the existing high-resolution samples \citep{origlia11}.

Taken together, Terzan~5 and Liller~1 suggest useful BFF diagnostics rather than a rigid definition.  Their broad metallicity distributions are built from populations with distinct enrichment histories: old, metal-poor stars are $\upalpha$-enhanced, while later metal-rich populations record Type~Ia enrichment, and the available spectroscopy lacks the classical globular-cluster light-element pattern.  The relevant comparison is therefore population coherence rather than metallicity breadth alone.  An instructive analogue is $\omega$~Cen: \citet{carretta10} showed that combining M~54 with the surrounding Sagittarius nucleus produces a metallicity distribution and chemical pattern resembling $\omega$~Cen, consistent with composite nuclear systems observed at different evolutionary stages.

\section{The field bulge as the control population}

The BFF comparison is only useful if the ordinary bulge field is treated as a control population.  The Blanco DECam Bulge Survey (BDBS) provides an unusually large baseline for this purpose.  BDBS established a tight photometric relation between dereddened $(u-i)_0$ colour and spectroscopic [Fe/H] for red-clump giants, with a scatter of order 0.2 dex \citep{johnson20}; Fig 1.  Applying this relation to 2.6 million red-clump stars showed that the bulge metallicity distribution is intrinsically broad, asymmetric, and strongly dependent on Galactic latitude \citep{johnson22} (figs 1,2).

BDBS resolves metal-poor and metal-rich concentrations near [Fe/H]$\sim-0.3$ and $+0.2$ dex, respectively, but both distributions possess asymmetric tails and their relative importance changes strongly with Galactic latitude.  The metal-rich population is strongly concentrated toward the plane, whereas the metal-poor distribution becomes increasingly important at larger $|b|$ and shifts to lower metallicity \citep{johnson22}.  The survey therefore cautions against interpreting every peak in an observed metallicity distribution as a discrete stellar population: proximity to a metallicity peak alone does not establish common origin.

This point changes the most direct version of the BFF test.  The NSC cannot be identified as a fossil fragment merely because it has a wide metallicity range or a seemingly multi-peaked MDF.  The field bulge itself already has a complex MDF.  The stronger requirement is that candidate NSC components be narrower than expected from the field distribution and remain coherent in independent dimensions such as age, detailed chemistry, spatial distribution, or kinematics.

\begin{figure}
    \begin{center}
    \includegraphics[width=0.92\textwidth]{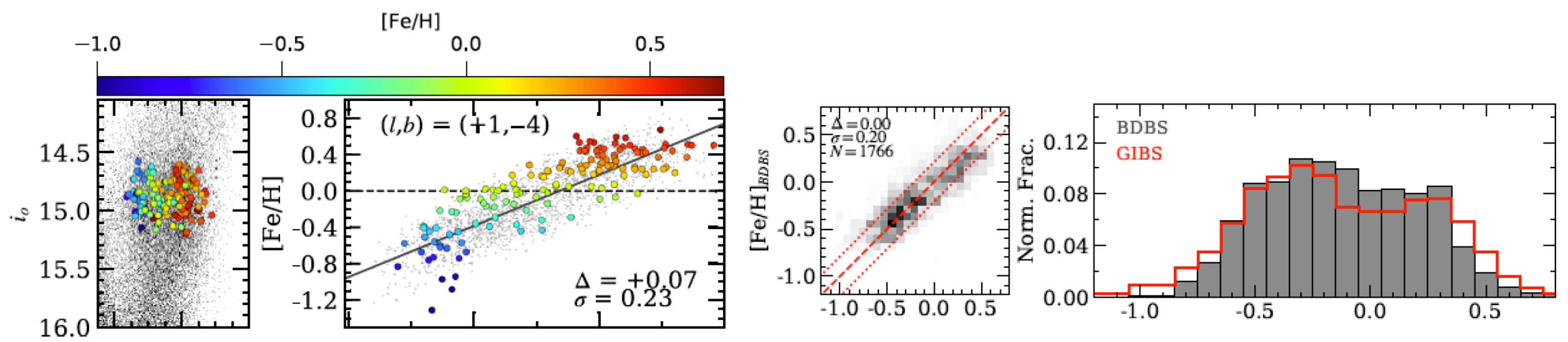}
    \caption{Photometric metallicities in the Blanco DECam Bulge Survey.  Left panel: the dereddened $(u-i)_0$ color (x-axis) of bulge red-clump giants provides a metallicity estimator with a scatter of approximately 0.2 dex when calibrated against [Fe/H] from high-resolution spectroscopy \citep{johnson20}.  Right: comparison of BDBS photometric metallicities with a large spectroscopic validation sample and the resulting metallicity distribution.  The BDBS calibration enables metallicity measurements for 2.6 million red-clump stars across the southern bulge \citep{johnson22}.}
    \label{fig:bdbs-calibration}
    \end{center}
\end{figure}

\begin{figure}
    \begin{center}
    \includegraphics[width=0.92\textwidth]{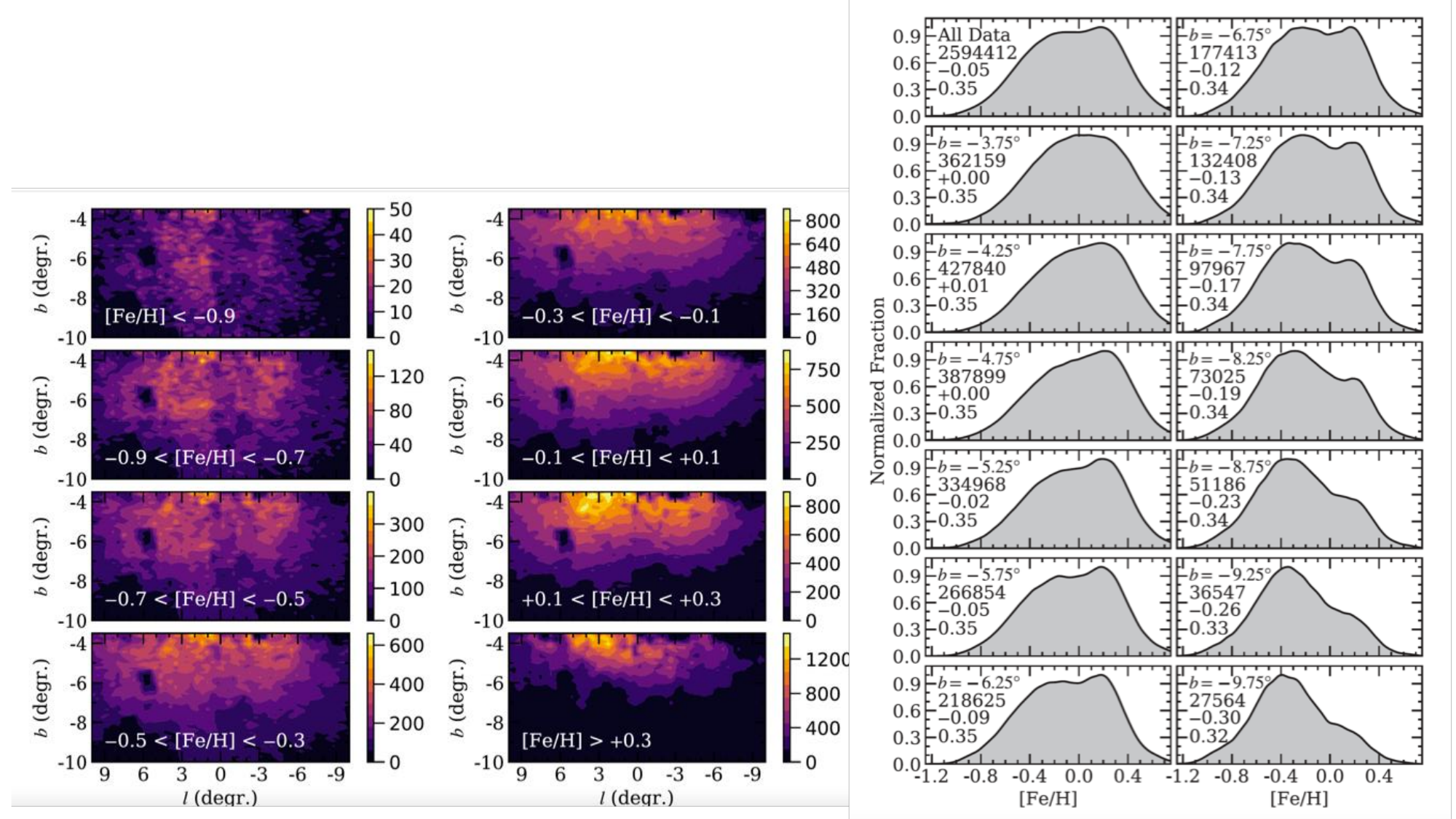}
    \caption{Bulge metallicity structure from BDBS \citep{johnson22}.  The metal-rich population is strongly concentrated toward the Galactic plane, while the shape of the metallicity distribution changes substantially with latitude.  The broad, asymmetric field-bulge MDF is an essential control when asking whether apparent metallicity components in the NSC represent discrete fossil populations.}
    \label{fig:bdbs-structure}
    \end{center}
\end{figure}

\section{Where does the nuclear star cluster fall?}

The first high-resolution metallicity survey of old NSC M giants already established an important connection: the NSC abundance distribution strongly resembled that of the Galactic bulge rather than the disc or halo \citep{rich17}.  This is qualitatively consistent with the idea that the nucleus contains material drawn from the same rapidly enriched inner-Galaxy environment that produced the bulge.  It is not, however, by itself evidence for a BFF, because the BDBS field population spans a similarly broad metallicity domain.

Detailed abundance ratios provide a more discriminating comparison.  At subsolar metallicity, NSC stars broadly overlap the inner-bulge $\upalpha$ sequence.  At supersolar metallicity, however, the Keck/NIRSPEC sample contains stars with enhanced [Si/Fe] relative to local thin-disc stars \citep{thorsbro20}.  This behaviour is interesting in the BFF context because the metal-rich population of Terzan~5 instead has nearly solar [$\upalpha$/Fe] \citep{origlia11}.  The simplest one-to-one mapping --- old metal-poor $\upalpha$-enhanced stars plus a younger, metal-rich, solar-$\upalpha$ population --- therefore does not reproduce the full NSC abundance pattern.  Recent R=40,000 infrared spectroscopy using the Immersion Grating Infrared Spectrograph (IGRINS;\cite{mace2016}) finds a declining $\upalpha$ trend at high metallicity and emphasizes similarities with the inner bulge, while also demonstrating that current high-resolution samples do not yet define a completely homogeneous metal-rich sequence \citep{ryde25}.  The NSC may contain real chemical substructure, but larger samples analyzed on a {\it common scale} are required to separate that signal from selection and analysis systematics.

The central parsec adds another level of complexity.  A companion contribution in this volume by Thorsbro and collaborators discusses high-resolution abundance measurements of stars projected close to Sgr~A*.  Those data include an old, subsolar-metallicity star with low [$\upalpha$/Fe] as well as metal-rich stars that may remain $\upalpha$-enhanced.  Such heterogeneity is naturally difficult to fit into a single closed enrichment sequence.  It is instead compatible with a nucleus that combines stars formed in situ from rapidly enriched gas with populations transported inward from the inner bulge, nuclear stellar disc, disrupted clusters, or other early building blocks.  Bar-driven inflow can feed the Central Molecular Zone, but transport onward into the central tens of parsecs is model-dependent and may require additional processes beyond the large-scale bar flow \citep{tress20}.

The light elements offer a particularly clean future discriminator.  In Liller~1 and Terzan~5, the absence of a Na--O or related light-element anticorrelation argues against a conventional globular-cluster origin \citep{origlia11,alvarez24}.  The corresponding experiment has not yet been carried out for a sufficiently large, homogeneous NSC sample.  Detecting a substantial population with genuine globular-cluster light-element anomalies would directly support an important cluster-infall contribution; their absence would make the old NSC chemically more BFF- or bulge-like.

\section{A falsifiable BFF hypothesis}

Table~\ref{tab:tests} summarises the main diagnostics.  No single observable is decisive.  The strength of the comparison comes from requiring the same stars to remain grouped in several dimensions at once.

\begin{table}
\caption{Observational tests of a BFF-like component in the nuclear star cluster.}
\label{tab:tests}
\begin{center}
{\scriptsize
\begin{tabular}{lll}
\hline
Diagnostic & BFF benchmark & Discriminating NSC test \\
\hline
MDF structure & Distinct [Fe/H] groups & Intrinsically narrow groups? \\
Age--metallicity & Distinct formation episodes & Ages of chemical groups \\
{[$\upalpha$/Fe]} trend & Old enhanced; young near solar & Coherent metal-rich enhancement? \\
Light elements & No GC-like anticorrelation & Na--O/Al--O anomalies present? \\
Population structure & Spatial segregation & Spatial/kinematic coherence? \\
\hline
\end{tabular}}
\end{center}
\end{table}

Several outcomes would then have distinct interpretations.  A small number of narrow, age-coherent metallicity groups with bulge-like chemistry and no globular-cluster light-element anomalies would favour a substantial BFF-like component.  Multiple such groups with differing kinematics could instead point toward assembly from several massive inner-Galaxy fragments.  An old population rich in Na--O or related globular-cluster anomalies would favour a larger contribution from inspiralling genuine globular clusters.  Finally, a chemically continuous distribution following the bulge sequence, combined with age and kinematic gradients rather than discrete groups, would favour repeated in-situ star formation plus migration from the surrounding inner Galaxy.

The $\upalpha$-element test is already capable of ruling out the simplest version of the hypothesis.  If the supersolar NSC population were directly analogous to the younger metal-rich component in Terzan~5, approximately solar [$\upalpha$/Fe] would be expected.  The existence of supersolar-metallicity, Si-enhanced NSC stars \citep{thorsbro20} therefore demands either a different enrichment timescale, additional gas infall, or a mixture of populations.  The NSC may still contain an early BFF-like component, but the present-day cluster is unlikely to be a straightforward surviving copy of Terzan~5 or Liller~1.

\section{Summary and outlook}

The comparison among the field bulge, bulge fossil fragments, and the NSC provides a useful way to organise the rapidly expanding Galactic-centre abundance data.  The BDBS field-bulge sample shows that a broad or even structured metallicity distribution is not by itself evidence for discrete fossil populations \citep{johnson22}.  Terzan~5 and Liller~1 demonstrate what stronger evidence looks like: chemically and chronologically distinct populations, bulge-like enrichment patterns, and an absence of the light-element anticorrelations that characterise genuine globular clusters \citep{origlia11,ferraro21,alvarez24,zullo26}.

The NSC shares important properties with both comparison classes.  Its broad metallicity distribution and detailed abundances link it strongly to the inner bulge \citep{rich17,ryde25}, while its extended star-formation history and chemical heterogeneity resemble the complexity that motivated the BFF interpretation of Terzan~5 and Liller~1.  At the same time, the metal-rich $\upalpha$-enhanced population found by \citet{thorsbro20} shows that the analogy cannot be exact.

The decisive next step is therefore not simply more metallicities, but multi-dimensional chemical tagging.  High-resolution infrared spectra should identify candidate NSC components in [Fe/H], $\upalpha$ elements, and light elements, followed by age, spatial, and kinematic tests of those same groups.  The ELT may eventually test for multiple old main-sequence turnoffs in the NSC, analogous to those resolved in Terzan~5 and Liller~1.  In this form the BFF hypothesis is falsifiable: it predicts coherent stellar populations rather than merely a broad MDF.  Whether or not the NSC ultimately qualifies as a bulge fossil fragment, these tests will quantify how much of the nucleus was inherited from early bulge building blocks, how much arrived through cluster infall, and how much formed in situ at the Galactic centre.

\begin{acknowledgements}
The authors thank the organisers of IAUS~405 for the opportunity to present this work.  RMR acknowledges financial support from his late father Jay Baum Rich. RMR acknowledges support from grant JWST-GO-5502.  BT acknowledges financial support from the Wenner-Gren Foundation (WGF2022-0041).
\end{acknowledgements}

\bibliographystyle{iaulike_iau_short}
\bibliography{rich_iau405_references}

@ARTICLE{alvarez24,
  author = {{Alvarez Garay}, D.~A. and {Fanelli}, C. and {Origlia}, L. and {Pallanca}, C. and {Mucciarelli}, A. and {Chiappino}, L. and {Crociati}, C. and {Lanzoni}, B. and {Ferraro}, F.~R. and {Rich}, R.~M. and {Dalessandro}, E.},
  title = {{X-shooter spectroscopy of Liller 1 giant stars}},
  journal = {A\&A},
  year = {2024},
  volume = {686},
  pages = {A198},
  doi = {10.1051/0004-6361/202449595}
}

@ARTICLE{ferraro16,
  author = {{Ferraro}, F.~R. and {Massari}, D. and {Dalessandro}, E. and {Lanzoni}, B. and {Origlia}, L. and {Rich}, R.~M. and {Mucciarelli}, A.},
  title = {{The Age of the Young Bulge-like Population in the Stellar System Terzan 5: Linking the Galactic Bulge to the High-z Universe}},
  journal = {ApJ},
  year = {2016},
  volume = {828},
  pages = {75},
  doi = {10.3847/0004-637X/828/2/75}
}

@ARTICLE{ferraro21,
  author = {{Ferraro}, F.~R. and {Pallanca}, C. and {Lanzoni}, B. and {Crociati}, C. and {Dalessandro}, E. and {Origlia}, L. and {Rich}, R.~M. and others},
  title = {{A new class of fossil fragments from the hierarchical assembly of the Galactic bulge}},
  journal = {Nature Astronomy},
  year = {2021},
  volume = {5},
  pages = {311--318},
  doi = {10.1038/s41550-020-01267-y}
}

@ARTICLE{johnson20,
  author = {{Johnson}, C.~I. and {Rich}, R.~M. and {Young}, M.~D. and {Simion}, I.~T. and {Clarkson}, W.~I. and {Pilachowski}, C.~A. and {Michael}, S. and {Kunder}, A. and {Koch}, A. and {Vivas}, A.~K.},
  title = {{Blanco DECam Bulge Survey (BDBS) II: project performance, data analysis, and early science results}},
  journal = {MNRAS},
  year = {2020},
  volume = {499},
  pages = {2357--2379},
  doi = {10.1093/mnras/staa2393}
}

@ARTICLE{johnson22,
  author = {{Johnson}, C.~I. and {Rich}, R.~M. and {Simion}, I.~T. and {Young}, M.~D. and {Clarkson}, W.~I. and {Pilachowski}, C.~A. and {Michael}, S. and {Marchetti}, T. and {Soto}, M. and {Kunder}, A. and others},
  title = {{Blanco DECam Bulge Survey (BDBS) IV: Metallicity distributions and bulge structure from 2.6 million red clump stars}},
  journal = {MNRAS},
  year = {2022},
  volume = {515},
  pages = {1469--1491},
  doi = {10.1093/mnras/stac1840}
}

@ARTICLE{massari14,
  author = {{Massari}, D. and {Mucciarelli}, A. and {Ferraro}, F.~R. and {Origlia}, L. and {Rich}, R.~M. and {Lanzoni}, B. and {Dalessandro}, E. and {Valenti}, E. and {Ibata}, R. and {Lovisi}, L. and {Bellazzini}, M. and {Reitzel}, D.},
  title = {{Ceci n'est pas a globular cluster: the metallicity distribution of the stellar system Terzan 5}},
  journal = {ApJ},
  year = {2014},
  volume = {795},
  pages = {22},
  doi = {10.1088/0004-637X/795/1/22}
}

@INPROCEEDINGS{mace2016,
       author = {{Mace}, Gregory and {Kim}, Hwihyun and {Jaffe}, Daniel T. and {Park}, Chan and {Lee}, Jae-Joon and {Kaplan}, Kyle and {Yu}, Young Sam and {Yuk}, In-Soo and {Chun}, Moo-Young and {Pak}, Soojong and {Kim}, Kang-Min and {Lee}, Jeong-Eun and {Sneden}, Christopher A. and {Afsar}, Melike and {Pavel}, Michael D. and {Lee}, Hanshin and {Oh}, Heeyoung and {Jeong}, Ueejeong and {Park}, Sunkyung and {Kidder}, Benjamin and {Lee}, Hye-In and {Nguyen Le}, Huynh Anh and {McLane}, Jacob and {Gully-Santiago}, Michael and {Oh}, Jae Sok and {Lee}, Sungho and {Hwang}, Narae and {Park}, Byeong-Gon},
        title = "{300 nights of science with IGRINS at McDonald Observatory}",
    booktitle = {Ground-based and Airborne Instrumentation for Astronomy VI},
         year = 2016,
       editor = {{Evans}, Christopher J. and {Simard}, Luc and {Takami}, Hideki},
       series = {Society of Photo-Optical Instrumentation Engineers (SPIE) Conference Series},
       volume = {9908},
        month = aug,
          eid = {99080C},
        pages = {99080C},
          doi = {10.1117/12.2232780},
       adsurl = {https://ui.adsabs.harvard.edu/abs/2016SPIE.9908E..0CM}
}

@ARTICLE{nandakumar25,
  author = {{Nandakumar}, G. and {Ryde}, N. and {Schultheis}, M. and {Rich}, R.~M. and {Di Matteo}, P. and {Thorsbro}, B. and {Mace}, G.},
  title = {{The First Chemical Census of the Milky Way's Nuclear Star Cluster}},
  journal = {ApJL},
  year = {2025},
  volume = {982},
  pages = {L14}
}

@ARTICLE{origlia11,
  author = {{Origlia}, L. and {Rich}, R.~M. and {Ferraro}, F.~R. and {Lanzoni}, B. and {Bellazzini}, M. and {Dalessandro}, E. and {Mucciarelli}, A. and {Valenti}, E. and {Beccari}, G.},
  title = {{Spectroscopy Unveils the Complex Nature of Terzan 5}},
  journal = {ApJ},
  year = {2011},
  volume = {726},
  pages = {L20--L24},
  doi = {10.1088/2041-8205/726/2/L20}
}

@ARTICLE{pfuhl11,
  author = {{Pfuhl}, O. and {Fritz}, T.~K. and {Zilka}, M. and {Maness}, H. and {Eisenhauer}, F. and {Genzel}, R. and {Gillessen}, S. and {Ott}, T. and {Dodds-Eden}, K. and {Sternberg}, A.},
  title = {{The Star Formation History of the Milky Way's Nuclear Star Cluster}},
  journal = {ApJ},
  year = {2011},
  volume = {741},
  pages = {108}
}

@ARTICLE{rich17,
  author = {{Rich}, R.~M. and {Ryde}, N. and {Thorsbro}, B. and {Fritz}, T.~K. and {Schultheis}, M. and {Origlia}, L. and {J{\"o}nsson}, H.},
  title = {{Detailed Abundances for the Old Population near the Galactic Center. I. Metallicity Distribution of the Nuclear Star Cluster}},
  journal = {AJ},
  year = {2017},
  volume = {154},
  pages = {239},
  doi = {10.3847/1538-3881/aa970a}
}

@ARTICLE{ryde25,
  author = {{Ryde}, N. and {Nandakumar}, G. and {Schultheis}, M. and {Kordopatis}, G. and {Di Matteo}, P. and {Haywood}, M. and {Sch{\"o}del}, R. and {Nogueras-Lara}, F. and {Rich}, R.~M. and {Thorsbro}, B. and others},
  title = {{Chemical Abundances in the Nuclear Star Cluster of the Milky Way: {$\alpha$}-Element Trends and Their Similarities with the Inner Bulge}},
  journal = {ApJ},
  year = {2025},
  volume = {979},
  pages = {174}
}

@ARTICLE{thorsbro20,
  author = {{Thorsbro}, B. and {Ryde}, N. and {Rich}, R.~M. and {Schultheis}, M. and {Renaud}, F. and {Spitoni}, E. and {Fritz}, T.~K. and {Mastrobuono-Battisti}, A. and {Origlia}, L. and {Matteucci}, F. and {Sch{\"o}del}, R.},
  title = {{Detailed abundances in the Galactic center: Evidence of a metal-rich alpha-enhanced stellar population}},
  journal = {ApJ},
  year = {2020},
  volume = {894},
  pages = {26}
}

@ARTICLE{zullo26,
  author = {{Zullo}, G. and {Pallanca}, C. and {Ferraro}, F.~R. and {Lanzoni}, B. and {Origlia}, L. and {Massari}, D. and {Dalessandro}, E. and {Fanelli}, C. and {Cadelano}, M. and {Vesperini}, E. and {Crociati}, C. and {Rich}, R.~M. and {Valenti}, E.},
  title = {{The multi-age stellar populations of Terzan 5 as revealed by JWST}},
  journal = {A\&A},
  year = {2026},
  volume = {709},
  pages = {A212},
  doi = {10.1051/0004-6361/202659349}
}

@ARTICLE{carretta10,
  author = {{Carretta}, E. and {Bragaglia}, A. and {Gratton}, R.~G. and {Lucatello}, S. and {Bellazzini}, M. and {Catanzaro}, G. and {Leone}, F. and {Momany}, Y. and {Piotto}, G. and {D'Orazi}, V.},
  title = {{M 54 + Sagittarius = $\omega$ Centauri}},
  journal = {ApJL},
  year = {2010},
  volume = {714},
  pages = {L7--L11},
  doi = {10.1088/2041-8205/714/1/L7}
}

@ARTICLE{tress20,
  author = {{Tress}, R.~G. and {Sormani}, M.~C. and {Glover}, S.~C.~O. and {Klessen}, R.~S. and {Battersby}, C.~D. and {Clark}, P.~C. and {Hatchfield}, H.~P. and {Smith}, R.~J.},
  title = {{Simulations of the Milky Way's central molecular zone -- I. Gas dynamics}},
  journal = {MNRAS},
  year = {2020},
  volume = {499},
  pages = {4455--4478},
  doi = {10.1093/mnras/staa3120}
}

\end{document}